\documentclass[runningheads]{llncs}
\usepackage[T1]{fontenc}
\usepackage{xcolor} 
\usepackage{graphicx}
\usepackage{amsfonts}
\usepackage{amsmath}
\usepackage{enumitem}

\usepackage{booktabs}
\usepackage{tabularx}
\usepackage{booktabs}
\usepackage{array}
\usepackage{makecell}
\usepackage{arydshln}
\usepackage{caption}

\begin{document}
\title{The Effect of Network Topology on the Equilibria of Influence-Opinion Games}
\titlerunning{Network Topology and Influence-Opinion Games}
\author{Yigit Ege Bayiz\orcidID{0000-0002-0700-2768} \and
Arash Amini\orcidID{0000-0002-3713-8674} \and
Radu Marculescu\orcidID{0000-0003-1826-7646} \and
Ufuk Topcu\orcidID{0000-0003-0819-9985}}
\authorrunning{YE. Bayiz et al.}
%
\institute{The University of Texas at Austin, Austin TX 78712, USA
}
\maketitle              
\begin{abstract}

Online social networks exert a powerful influence on public opinion. Adversaries weaponize these networks to manipulate discourse, underscoring the need for more resilient social networks. To this end, we investigate the impact of network connectivity on Stackelberg equilibria in a two-player game to shape public opinion. 
We model opinion evolution as a repeated competitive influence-propagation process. Players iteratively inject \textit{messages} that diffuse until reaching a steady state, modeling the dispersion of two competing messages.
Opinions then update according to the discounted sum of exposure to the messages. This bi-level model captures viral-media correlation effects omitted by standard opinion-dynamics models. To solve the resulting high-dimensional game, we propose a scalable, iterative algorithm based on linear-quadratic regulators that approximates local feedback Stackelberg strategies for players with limited cognition. We analyze how the network topology shapes equilibrium outcomes through experiments on synthetic networks and real Facebook data. Our results identify structural characteristics that improve a network's resilience to adversarial influence, guiding the design of more resilient social networks.

\end{abstract}
\section{Introduction}
\vspace{-5pt}

Withstanding adversarial influence in social networks is a pressing issue across domains ranging from marketing to online influence campaigns\cite{truong2024quantifying,hajaj2021robust}. In such settings, two competing players often engage in information conflict, each injecting messages into the network to influence public opinion. One player, the \textit{adversary}, strategically spreads a harmful or misleading message, while the other, the \textit{defender}, counters with a corrective or opposing message. This competition unfolds over time as each player sequentially inject messages into the network, aiming to influence the distribution of public opinion in their favor. 
Social networks evolve continually; as public opinion changes, individual connections adapt accordingly\cite{gu2017co,kandel2017homophily}. Consequently, to prevail in these information conflicts, players must not only maximize their immediate influence but also strategically shape the network---through targeted opinion distribution---to sustain their advantage over time.

Accurately modeling this competitive opinion evolution is therefore crucial for understanding network resilience against adversarial influence\cite{gaitonde2020adversarial,golpayegani2024adaptation,amini2024control}. 
However, existing models of opinion dynamics and influence propagation largely fail to capture this multi-round competition, often considering only a single contagion spreading in isolation \cite{kempe2003maximizing,borodin2013strategyproof}, or assuming static, non-adaptive influence sources\cite{zuo2022online}. In contrast, real-word influence conflicts involve co-evolving opinions and the game-theoretic interplay between the long-term decision-making of adversaries and defenders.

\textit{This paper} introduces a \textit{bi-level} competitive opinion update model to study conflicts in information environments, integrating a lower-level diffusion process for modeling competing influences and a top-level opinion update model. Formally, we consider a repeated sequential game between two players. At each round, the adversary (\textit{leader}) introduces an opinionated \textit{message} into the network, originating from a chosen point in the opinion space. The defender (\textit{follower}) then responds by injecting a competing message. 
These \textit{messages} then propagate through a competitive diffusion process. After diffusion, each individual in the network is predominantly influenced by either the defender or the adversary, reflecting the realistic constraint that individuals often choose between conflicting pieces of information. Once the influence diffusion process reaches an equilibrium state, the \textit{opinions} of the individual update based on which player has the predominant influence over the individual. A key advantage of this bi-level approach is its ability to capture correlation effects message exposure and opinion evolution that are often missed by standard opinion dynamics models, such as the Friedkin-Johnsen type \cite{friedkin1990social,jia2015opinion}. These standard models typically assume that an individual's opinion is simultaneously affected by all neighbors---subject to connection weights---irrespective of the specific influence each neighbor propagates at a given time. We assume that each player makes decisions according to the Stackelberg feedback equilibrium \cite{simaan1973stackelberg}.

In summary, our key contributions are threefold. First, we introduce a novel bi-level influence-opinion model that formalizes the long-term opinion effects arising from the interplay between a malicious influence source and a defensive counter-influence source. Second, we formulate this attacker-defender interaction as a Stackelberg game and derive scalable, approximate local-feedback Stackelberg equilibrium solutions. Finally, we investigate how the homophily of social networks influences the resilience of public opinion to adversarial influence.

\vspace{-10pt}
\section{Problem Statement}

Adversaries are increasingly exploiting social networks, creating a competitive dynamic in which defenders must counteract adversarial influence aimed at manipulating public opinion \cite{stella2018bots,zannettou2020characterizing,budak2011limiting}. However, existing models often overlook the long-term nature of these campaigns and the co-evolution of beliefs and network, typically focusing on short-term influence spread or isolated dynamics \cite{kempe2003maximizing,zuo2022online,bayiz2024optimization}. To bridge this gap, we model the adversarial competition as a bi-level influence-opinion Stackelberg game, where the adversary acts as the leader. This paradigm uniquely integrates rapid message propagation (influence spread) with the slower-paced evolution of public opinions and social ties, leveraging time-scale separation to better understand how persistent exposure shapes long-term beliefs and ultimately, network resilience against such manipulation.

Consider a network of $n$ individuals that interact according to the time-varying communication graph $\mathcal{G}_t=\{\,V\,,\,W_t\,\}$. Here, $V:=\{\,1\,,\, \ldots\,,\, n\,\}$ represents the set of individuals, and $W_t \, \in \,[\,0\,,\,1\,]^{n \times n}$ stores the probabilities of interaction between individuals at time $t$. Each individual $i\,\in \,V$ has an opinion $x^i_t \,\in \,\mathbb{R}^d$ in macro-time $t$. We assume that the underlying interactions are correlated with the individual's opinion. Specifically, the probability of individuals $i,j \,\in\, V$ interacting is correlated by $\psi(\,x^i_t\,,\,x^j_t\,)$, where $\psi(\,\cdot\,,\,\cdot\,): \mathbb{R}^d \times \mathbb{R}^d \rightarrow [\,0\,,\,1\,]$ is the interaction kernel. The $[W_t]_{ij}$ is the row normalization of the matrix of interaction kernels, i.e., 
\begin{equation}\label{weight.matrix}
    [W_t]_{ij} \, = \,\frac{\psi(\,x^i_t\,,\,x^j_t)\,}{\sum_{k\in V/{i}} \,\psi(\,x^i_t\,,\,x^j_t\,)}, ~\forall ~i\neq j\, \in \,V,
\end{equation}
and $[W_t]_{ii}\,=\,0$. In each macro-time step $t$, the adversary initiates by sharing messages that support the opinion $u^a_t \, \in \, \mathbb{R}^d$. This is equivalent to strategically spreading messages from network nodes with opinions similar to $u^a_t$, using homophily-based connections. The defender then responds by choosing a message with opinion $u^d_t \, \in \, \mathbb{R}^d$ to counter the adversarial messages. We assume that the adversary aims to shift the opinions of individuals toward a target state $x_a$ (known to both players), while the defender strives to maintain the status quo. Following initial message selections, individuals spread information rapidly across the network through the communication graph $\mathcal{G}_t$.  Given that messages propagate significantly faster than opinions evolve, we model information propagation by linear dynamics, with another time scale, $s > 0$, denoted by micro-time. 

The message of the defender aims to counter the adversarial influence. Therefore, we assign an \textit{evidential value} of $1$ to the defender's message and $-1$ to the adversary's message. In this way, we can abstract the propagation of opposing information. If an individual simultaneously receives adversarial and defender content, their evidential values cancel each other out. Let us assume that at some micro-time $s$, individual $i$ has already accumulated evidential information $y^{i}_{s}\, \in \, \mathbb{R}$. Individuals share their current evidential information with a probability $\alpha$, and the evidential information for the individual $i$ evolves by
\begin{equation}
    y_{s+1} \,=\, \alpha W_t y_{s} \,-\, p^{a}_t e^{-\kappa_a s} \,+\, p^{d}_t e^{-\kappa_d s}
\end{equation}
Here, $[p^\square_t]_i = \psi(\,u^\square_t\,,\,x^i_t\,)$ represents the probability that $i$ observes message shared by the adversary(or defender), based on their distance of opinions, $\square = a$(or $\square = d$). We incorporate an exponential decay for the probability of interaction to model the community's diminishing interest in the messages over time. To maintain heterogeneity, we assume that the rate of interest decay differs for adversarial ($\kappa_a$) and defender ($\kappa_d$) messages.

We calculate the overall evidential information each individual receives during these interactions by
\begin{equation}
    \bar y_t \,:= \,\sum_{s=0}^\infty \,y_s =\, (I-\alpha W_t)^{-1} \,\Big[\,p^d_t \frac{e^{-\kappa_d}}{1- e^{-\kappa_d}} \,-\, p^a_t \frac{e^{-\kappa_a}}{1- e^{-\kappa_a}} \,\Big].
\end{equation}
Here, $\bar y$ represents the total accumulated evidence for each individual. Individuals then shift their opinions towards the defender's content $u_t^d$ with a probability $\varsigma(\bar y^i)$, where $\varsigma(\cdot):\mathbb R \rightarrow [0,1]$ is a sigmoid function, and towards the attacker's content $u^a_t$ with a probability $1-\varsigma(\bar y^i)$. The individual opinions evolevs by
\begin{equation} \label{eq:OpDyn}
    x^i_{t+1}\, = \, (1-\lambda) x^i_0 \,+ \, \lambda\Big(\,x_t +\eta\,\rvert \bar y^i_t \lvert\,\big( \,\varsigma(\bar y^i_t)\,u^d_t\, + \,(\,1-\varsigma(\bar y^i_t)\,)\,u^a_t -x_t\,) \big)\Big),
\end{equation}
where $x^i_0$ denotes the initial opinion of individual $i$ at time $0$. This opinion evolution model is derived from the Friedkin-Johnson opinion dynamics model, with a learning rate $\eta\rvert \bar y^i_t \lvert$.  In this model, individuals exhibit a degree of stubbornness toward their initial opinions, represented by $\lambda$. Note that learning rate reflect the amount of evidential information individual receives.

We assume that both the adversary and the defender aim to minimize a quadratic cost function. We heavily penalize sharing extreme ideological content, especially for the defender, who needs to remain neutral. For a given adversary policy $\pi^a_t(x)$, the defender seeks to solve the  optimization problem
\begin{equation}
\begin{aligned}\label{defender.optim}
    \max_{u^d_t\sim\pi^d_{1:T}} \quad & J_d\,:= \,\sum_{t=1}^T\Big[\, (x_t-x_0)^TQ_d(x_t-x_0)\,+\,{u^{d}_t}^T R_d u^d_t)\,\Big]\\
    \textrm{s.t.} \quad & x_{t+1} = F(\,x_t\,,\,\pi^a_t(x_t)\,,\,\pi^d_t(x_t)\,),\\
\end{aligned}
\end{equation}
where $F(\cdot,\cdot,\cdot)$, summarize the evolution of dynamics for the individuals. Solving \eqref{defender.optim} yields the optimal defender policy $\pi^{*d}_t(x_t,\pi^a_t)$. The adversary then solves
\begin{equation}
    \begin{aligned}
        \max_{\pi^a_{1:T}} \quad & J_a\,:=\, \sum_{t=1}^T\Big[ \,(x_t-\hat x)^T Q_a (x_t- \hat x)\,+\, {u^a_t}^T R_a u^a_t)\, \Big]\\
        \textrm{s.t.} \quad & x_{t+1} = F(\,x_t\,,\,\pi^a_t(x_t)\,,\,\pi^{*d}_t(x_t,\pi^a_t)\,).\\
    \end{aligned}
\end{equation}

Social networks often involve millions of people within a vast opinion space. Even in focused scenarios where the opinion space is reduced to a few dimensions, the number of individuals can still be a few thousand. This immense state space presents a significant challenge in solving the described optimization problem. In the following section, we introduce a dynamic clustering-based approach designed to efficiently address this challenge.

\section{Methodology}
In this section, we describe a scalable method to approximately compute a local feedback Stackelberg equilibrium. Our approach relies on dynamic clustering of the population to reduce the dimensionality of the problem. We then use a model-predictive-control approach to find an approximate local feedback Stackelberg equilibrium for the problem.

\vspace{-15pt}
\subsubsection{Dynamic Clustering}
Real-world social networks often encompass millions of individuals, rendering the direct computation of local-feedback Stackelberg equilibria intractable due to their large scale. To address this curse of dimensionality, our model represents the individual population as a dynamic set of time-varying clusters, with assignments updated online. This approach enables operation on a \textit{reduced graph} of connected clusters, the order of which is significantly smaller than the node count of the original social network.

 In the initial macro-time step we construct the reduced graph by hierarchically clustering the network $\mathcal{G}$ into $m_0$, and building the quotient graph $\hat{\mathcal{G}}_0 = (\hat{\mathcal{S}}_0, \hat W_0)$, where $\hat{\mathcal{S}}_t = \{s_t^1, \dots\ s_t^{m_t}\}$ is the set of clusters in macro-time $t$. We then treat each cluster as an individual and find $W_t$ according to \eqref{weight.matrix}.

Rather than re-clustering at each time step, which can be computationally burdensome, we compute the cluster assignments $\hat{\mathcal{s}}_t$ in all subsequent time steps by sequentially applying splitting and merging procedures on the clusters $\hat{\mathcal{s}}_{t-1}$ in the previous time step. This process aims to split bimodal clusters into unimodal cluster pairs and merge pairs of clusters with unimodal mixture distributions into a single cluster.

\vspace{-13pt}
\paragraph{Cluster Splitting} The splitting procedure iteratively splits clusters with sufficiently high bimodality into individual clusters by running a hierarchical clustering algorithm over the clusters with a fixed target cluster size of $2$. We determine the splitting threshold for each cluster by tracking the skewness, $\gamma_1$, and kurtosis, $\gamma_2$, for the opinion distribution across the principal axis of each cluster and calculating the Sarle's bimodality coefficient, $\textrm{BC} = {(\gamma_1^2 + 1)}/{\gamma_2}.$ We then compare this coefficient to a threshold value and split the cluster if the bimodality coefficient exceeds the threshold value. For our experiments, we found that a threshold of $0.55$ works well and leads to stable cluster counts.

\vspace{-13pt}
\paragraph{Cluster Merging} The merging rule we employ aims to ensure that all clusters $S_t^i$ in the network have unimodal distributions. For each cluster $i$, we first compute the mean $\mu_i$ and the covariance matrix $\Sigma_i$. We then iterate over cluster pairs and greedily merge each pair if the means of each cluster in the pair lie within one standard deviation away from the other cluster's mean opinion. That is, we merge clusters $i, j$ whenever the following condition holds for $d = \mu_i - \mu_j$,
\begin{equation}
    ||d|| ^2 < \min \big(\frac{d^\top\Sigma_i d}{d^\top d},\; \frac{d^\top\Sigma_j d}{d^\top d}\big).
\end{equation}

The above is a sufficient condition that guarantees the mixture distribution of the clusters $i$ and $j$ is unimodal, given that clusters $i$ and $j$ each have normal distributions \cite{Behboodian2012mixture}. Clearly, this assumption is not guaranteed to hold in the real-world networks; however, in practice, we observe that the competition between the players causes sufficient mixing for the cluster distributions to tend towards normal distributions. 

\vspace{-15pt}
\subsubsection{Bounded Cognition Stackelberg Equilibrium}


Recognizing that in practice decision-makers in this game are human experts whose behavior often deviates from the perfect rationality assumed in traditional game theory, our methodology departs from the perfect Stackelberg equilibrium concept \cite{rosas2010evolutionary,kurtz2024limits}. Instead, we address the complexities of human cognition alongside the inherent high-dimensionality and nonlinearity of the problem by employing an iterative approach. In this approach, players model their opponent as possessing a cognitive capacity one level lower. To effectively manage persistent non-linearity, we first approximate the evolution of opinion cluster centers via linearization. Subsequently, we determine the feedback policy for each player by sequentially solving the resultant time-varying linear quadratic regulator (LQR) problem, based on the assumption that the opposing player operates according to this one-level-lower cognitive model \cite{vamvoudakis2022nonequilibrium}. We implement a receding-horizon control approach to model the decision making of the players. We assumes that players update their policies and cluster formations by observing the actual evolution of the social network after a period of policy application. This strategy is particularly well-suited for the inherent unpredictability and high uncertainty of social network dynamics, where reliable predictions are typically confined to short horizons.

Let $s_t \in \mathbb R^m_t$ denote the position of the centers of the clusters in some macro-time $t$, known to both players.
Assume that we have calculated a reference trajectory, $\hat s_{t:t+H}^{0}$ as an initial estimate of the trajectory for the next $H$ steps, calculated by assuming both the adversary and the defender play fixed policy  $u^{a,0}_{t:t+h} = u^a_{t-1}$ and $u^{d,0}_{t:t+h} = u^d_{t-1}$ over the horizon.
A first-order expansion of dynamics around the reference trajectory gives the perturbation dynamics
\begin{equation}
  \bar s^0_{t+1}=A_t^0\, \bar s_t+B_t^{a,0}\, u_t^{a}+B_t^{d,0}\, u_t^{d},
  \tag{10}
\end{equation}
\[
  A_t^0=\left.\frac{\partial F}{\partial s}\right|_{(\hat s_t^0,\hat u_t^{a,0},\hat u_t^{d,0})},
  \qquad
  B_t^{\square}=\left.\frac{\partial F}{\partial u^{\square}}\right|_{(\hat s_t^0,\hat u_t^{a,0},\hat u_t^{d,0})},
  \quad\square\in\{a,d\}.
\]

Let us denote the feedback policy for the defender and attacker at the cognition level $\ell$ by $u^{d,\ell}_{t:t+H}(s),~u^{d,\ell}_{t:t+H}(s)$. Since the $t-1$ messages are available to both players, they both have an approximation of the other player zeroth cognition level input. At each cognition level the defender (\textit{follower}) needs to solves
\begin{equation}\label{defender_optim}
\begin{aligned}
    \min_{\{u_{\tau}^{d,l}\}_{\tau=t}^{t+H-1}}
  \sum_{\tau=t}^{t+H-1} & \Bigl[
       (\bar s_{\tau}^{\ell-1})^{T}\!Q_{d}\,\bar s_{\tau}^{\ell-1}
     + (u_{\tau}^{d,\ell-1})^T R_{d}\, u_{\tau}^{d,\ell-1}
  \Bigr]\\
  s.t.~~~~~~~~~~~~~ &   \bar s_{t+1}^{\ell-1}=(A_t^{\ell-1} + B_t^{a,\ell-1}\,K^{a,\ell-1}_t ) \bar s_t^{l-1}+B_t^{d,\ell-1}\, u_t^{d,\ell-1}.
\end{aligned}
\end{equation}
We can solve the optimization problem \eqref{defender_optim} by standard Riccati recursion to find the feedback for modified state matrix $\hat A_{t}^{d,l}=A_t^{\ell-1} + B_t^{a,\ell-1}\,K^{a,\ell-1}_t$.
\begin{equation}
    K_{t}^{d,l}=-(R_{d}+{B_t^{d,\ell-1}}^T\,P_{t+1}^{d,l}B_t^{d,\ell-1})^{-1}\, {B_t^{d,\ell-1}}^T P_{t+1}^{d,l} \hat A_{t}^{d,l},
\end{equation}
where $P_{t}^{d,l}$ satisfies
\[
  P_{t}^{d,l}=Q_d+{(\hat A_{t}^{d,l})}^{T}\!\Bigl(P_{t+1}^{d,l}-P_{t+1}^{d,l}B_t^{d,\ell-1} {S_{t}^{d,l}}^{-1}{\left(B_t^{d,\ell-1}\right)}^T P_{t+1}^{d,l}\Bigr){\hat A_{t}^{d,l}},
\]
\[
S_t^{d,l}=R_d+{B_t^{d,\ell-1}}^T P_{t+1}^{d,l}B_t^{d,\ell-1}.
\]

The Adversary (\textit{leader}) then consider the  $\ell$-th level response of the defender $ u^{d,l}_t(s)=  K_{t}^{d,l} \bar s_t^{\ell-1}$, and forms its optimization problem similar to \eqref{defender_optim}. Adversary (\textit{leader}) solves the optimization using standard Riccati recursion similar to the defender (\textit{follower}) using  the modified state matrix, $\hat A_{t}^{a,l}=A_t^{\ell-1} + B_t^{d,\ell-1}\,K^{d,\ell}_t$.

We then update the estimate of the trajectories, $\hat s_{t:t+H}^{\ell}$, according to the $\ell$-th level feedback policies. We repeat this approach until we reach the maximal cognitive level assumed for each player. After the last step, we solve the defender problem one more time, to maintain the Stackelberg hierarchy, assuming that defender observes the adversary policy, and find the best response to that policy. The players then apply the policies over short span of time, and then get exact updates from the social network. This model-predictive implementation captures the empirical reality that reliable forecasts on social platforms rarely extend beyond a few days.

\vspace{-10pt}
\section{Results and Discussion}
\vspace{-5pt}

\begin{table}[t]
    \centering
    \begin{tabularx}{310 pt}{ccccc}
        \Xhline{2\arrayrulewidth}
        Parameter & Description & Synthetic Network& & Facebook Network \\ 
         \hline
         $\lambda$ & Stubbornness & 0.70& & 0.70\\ 
         $\eta$ & Learning rate & 0.50& & 1 \\ 
         $\alpha$ & Sharing probability & 0.30  &  & 0.5 \\ 
         $\sigma$ & Homophily coefficient & 1 & & 0.32 \\ 
         $\bar \ell$ & Maximum cognition level &  10 & & 10 \\ 
         $n$ & Number of Individuals  &  3000 & & 4038 \\ 
         \Xhline{2\arrayrulewidth}
    \end{tabularx}
    \caption{Experiments parameters}\label{Table:params} \vspace{-20pt}
\end{table}

\begin{figure}[t]
    \centering
    \includegraphics[width=\linewidth]{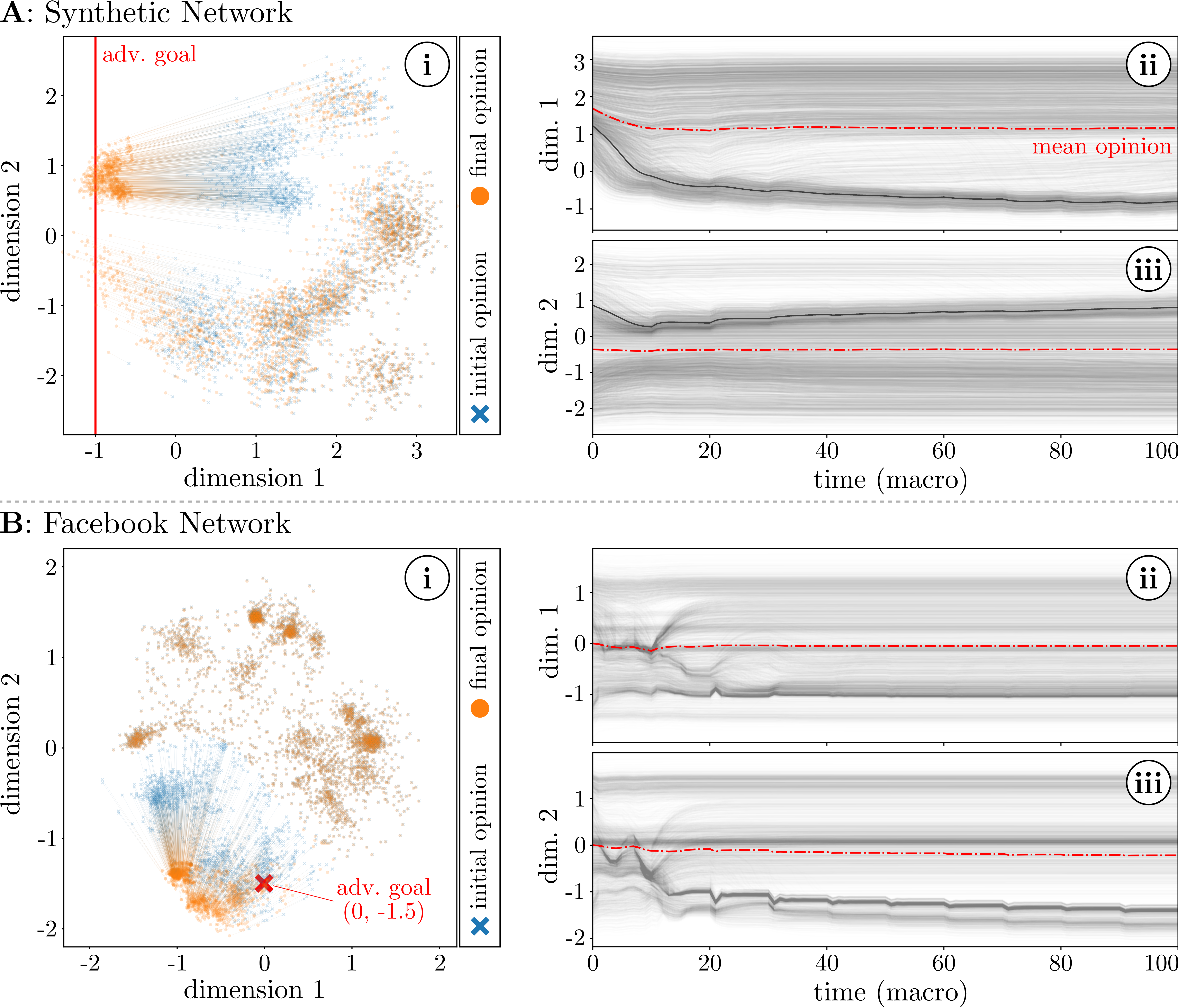}
    \caption{\textbf{Trajectories of individuals under competition.} Figures \textbf{A} and \textbf{B} show the results for synthetic networks and the Facebook network, respectively. From left to right, both panels display (i) the scatter plot of how each individual's opinion changes between initial and final state, and (ii, iii) the opinion changes of all individuals across macro-time steps for both principal dimensions. The dashed red lines show the average opinion of the population across time.
    }\vspace{-20pt}
    \label{fig:trajectories}
\end{figure}

We investigate the effect of network connectivity on a synthetic network constructed according to homophily interactions and a real network induced by Facebook data \cite{leskovec2012learning}. Table \ref{Table:params} provides the parameters we used for each network. We utilized the \textit{NVIDIA Ada Lovelace L4 Tensor Core GPU} for all experiments and the JAX package for automated derivatives to linearize dynamics and approximate equilibria. We assume a two-dimensional opinion space ($d=2$) for both cases. For the real network, we used the Fruchterman-Reingold force-directed algorithm to find $2$-D embeddings\footnote{All of the experiment codes used to generate the results in this section are accessible from \url{https://github.com/ege-bayiz/influence-opinion-games}}.

Figure \ref{fig:trajectories}-(A,B) illustrates (i) the evolution of the network and (ii-iii) the individual trajectories of the synthetic and Facebook networks, respectively. For the synthetic network, the adversary's cost function considers only the first dimension; that is, the adversary aims solely to shift the first dimension of individual opinions toward $-1$. In the Facebook network, the adversary targets the opinion point $x_a=[\,0\,,\,-1.5\,]$, which corresponds to an adversarial cost matrix $Q_a = 3I$. In both scenarios, we set the adversary's regulation cost $R_a=20I$. The defender has target and regulation costs of $Q_d=I$ and $R_d=80I$, respectively, modeling the fact that taking extreme actions is substantially more costly for the defender than for the adversary.

We observe that in both scenarios (Figure \ref{fig:trajectories}-(A,B)-i), the adversary targets nearby clusters and shifts their opinions toward its target. Especially in the synthetic network, the adversary manages to form its own echo chamber by exploiting the second dimension of the opinion space that it has freedom in (Figure \ref{fig:trajectories}-A-iii). By aligning the second dimension with a dense cluster, the adversary gains control of a large portion of the population and manages to drive the opinions of a few clusters toward its objective(Figure \ref{fig:trajectories}-A-i). We observe in Figure \ref{fig:trajectories}-(A,B)-ii that the equilibria result in an asymmetry among individuals; part of the network moves toward the adversary's target, while the defender manages to maintain the status quo for the remainder. This observation aligns with empirical results that show an asymmetric adverse influence \cite{williamson2016take,rao2022partisan}. Our results suggest that this asymmetry arises because the adversary exploits the underlying network topology and invests in specific parts of the network, rather than attempting to shift all individuals. Validating this observation and establishing a causal relation requires further theoretical and empirical experiments.

\begin{figure}[t]
    \centering
    \includegraphics[width=\linewidth]{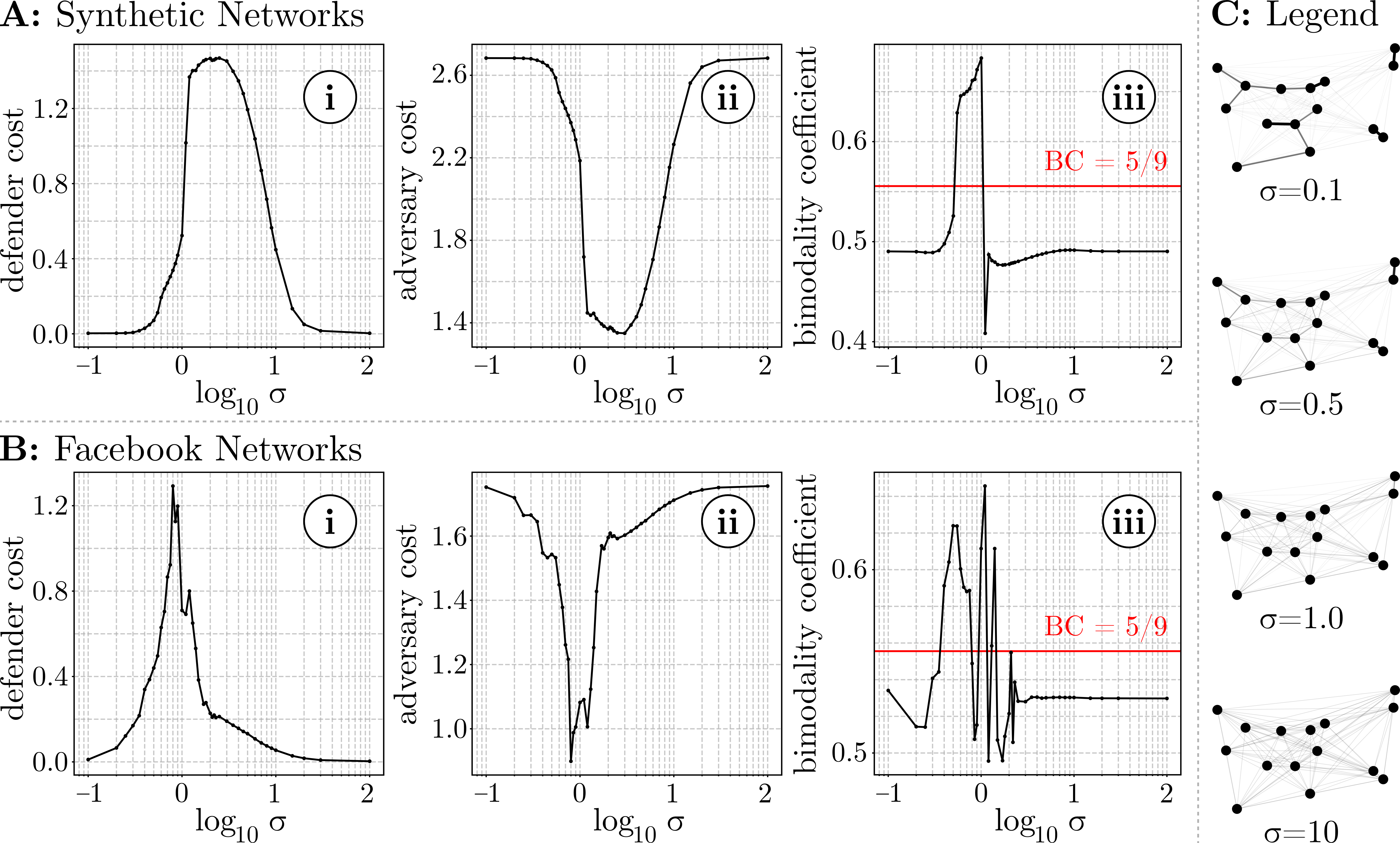}
    \caption{\textbf{Final state statistics across homophily kernels.} Figures \textbf{A} and \textbf{B} show the results for synthetic networks and the Facebook network, respectively. From left to right, both panels display (i) the mean distance to the defender's goal, (ii) the mean distance to the adversary's goal, and (iii) the bimodality coefficient of the final state $x_{T}$. Figure \textbf{C} shows small sample networks for different kernel parameters, illustrating how the network topology varies with the kernel parameter $\sigma$.}\vspace{-20pt}
    \label{fig:vs_sigma}
\end{figure}

Figure \ref{fig:vs_sigma}-(A,B) presents the sensitivity analysis of the defender and adversary costs with respect to the homophily coefficient, $\sigma$. In both scenarios, we observe a similar trend; low and high values of $\sigma$ make the network more resilient against the adversary. Low values of $\sigma$ indicate strong local connections and weak global connections, as demonstrated in Figure \ref{fig:vs_sigma}-C (for $\sigma = 0.1$). High locality hinders the adversary's ability to exploit network topology---low connectivity interrupts massage passing--- making the formation of a strong adversarial community difficult. In contrast, for large $\sigma$ values, the network approaches a complete graph with identical connections, as depicted in Figure \ref{fig:vs_sigma}-C (for $\sigma = 10$). In this scenario, regardless of the locations the adversary and defender choose to start their messages, the network's symmetry cancels their effects, thereby preventing the adversary from shifting opinions.

However, social networks typically operate with moderate $\sigma$ values; they are neither purely local, which would prevent the circulation of information, nor complete graphs, which would overwhelm users with unrelated information. Figure $\ref{fig:vs_sigma}$ shows that with moderate values of the homophily coefficient, the adversary gains an advantage and can shift a portion of the population towards its target. We emphasize that our approach approximates local linearized solutions; consequently, for the real network, we observe fluctuations in the costs of the defender and the adversary (Figure \ref{fig:vs_sigma}-B-(i,ii)).

Traditional approaches to counter adversarial influence often focus on network control, such as by removing edges or altering the information flow\cite{amini2024control,bayiz2024optimization}. However, our results demonstrate an important phenomenon. When dealing with a strategic adversary, understanding the current level of homophily is crucial, as adjusting it incorrectly can further exacerbate the problem and enable additional exploitation by the adversary.

Figure \ref{fig:vs_sigma}-(A,B)-iii demonstrates the bimodality coefficient, $\beta$, with respect to the homophily coefficient, $\sigma$. Our results reveal a phase transition for the synthetic network. As we increase $\sigma$, the bimodality coefficient initially increases, indicating that the adversary only captures a portion of the population. However, increasing $\sigma$ beyond a certain threshold changes the equilibrium behavior. In this phase, the adversary cam attracts most of the population, creating a uni-modal distribution around its target. As $\sigma$ increases further, the adversary gradually loses this advantage and the population shifts back to its initial state.

\vspace{-10pt}
\section{Conclusion}
\vspace{-5pt}

We study the effect of network topology to adversarial information operation in online social networks by introducing a novel bi-level influence-opinion game model. We formulate the problem as a Stackelberg game and present a scalable algorithm to approximate the equilibria for players with limited cognition. This approach captures correlation effects between the message propagation and the evolution of opinions alongside adaptive strategies that are often neglected in standard models. Our empirical investigations on synthetic and real-world Facebook networks revealed that network topology, particularly network connectivity, critically shapes adversarial resilience. We observe that both highly local and complete network topologies enhance network resilience, while moderate levels of connectivity can render networks vulnerable to adversarial influence, often leading to asymmetric opinion shifts, polarization, and favoring adversaries. These findings underscore the non-trivial impact of network topology and highlight that interventions aimed at altering network structure must be carefully considered, contributing to a deeper understanding of information conflict dynamics for designing more robust social platforms and effective counter-influence strategies.

\begin{credits}
\subsubsection{\ackname} We would like to thank Ashwin Ram for his discussions in the preliminary stages of problem formulation. 

This work was supported in parts by ARO under grant number W911NF-23-1-0317, by DARPA under grant number HR001123S0001, and by ONR under grant number N00014-22-1-2703.

\subsubsection{\discintname}
The authors have no competing interests to declare that are relevant to the content of this article.
\end{credits}
%
%
%
\bibliographystyle{splncs04}
\bibliography{Ref}

\noindent

\end{document}